# Aragats high-altitude research station – 80 years of continuous cosmic ray monitoring

*Z.Asaturyan\*, A.Chilingarian*

A I Alikhanyan National Laboratory, Cosmic Ray Division (CRD)
Alikhanyan Brorgers St. 2, Yerevan, Armenia, AM 0036

1. ## INTRODUCTION

The Aragats Cosmic Ray Research Station, established in 1943 amidst the challenges of World War II, has steadfastly advanced the study of particle and cosmic ray physics. The station is perched just beneath Armenia's majestic Mt. Aragats at 3200 m elevation under four prominent peaks. The northern summit, towering at 4090 m, overlooks a volcanic crater formed by an eruption 1.5 million years ago that covered half of Armenia with tuff and basalt. The station lies on a plateau adjacent to the frigid Kari Lake at the coordinates 40.4713N and 44.1819E.

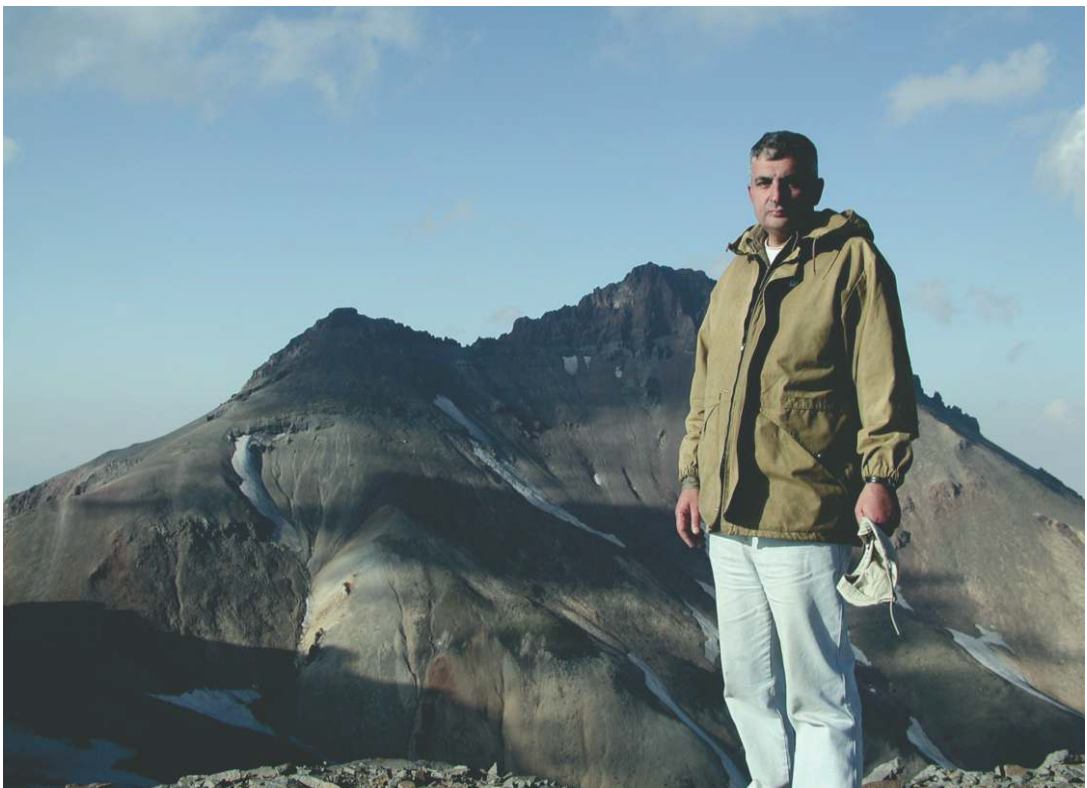

**Figure 1. A. Chilingarian on the Aragats south peak (3879 m). In the background, Aragats' highest North peak (4009 m), 1 September 2002. '**

*A. Chilingarian, the CRD head since 1992 and chairman of ASEC and ANI collabor,ations began his experimental work on Aragats in 1971.

Cosmic ray (CR) research on Aragats can be traced to 1934, with a study on East-West cosmic ray anisotropy by the expedition from the Leningrad Physical-Technical Institute, augmented by Norair Kocharian of Yerevan State University. The findings captivated Artem and Abraham Alikhanians, prompting an organized expedition to Aragats in 1942. Despite the adversities of global and local conflicts and scarcity of funds, electricity, and fuel that have marked Armenia's recent history, research expeditions have persevered for eight decades.

In its nascent stage, the exploration of cosmic rays opened a window to the universe's profound mysteries, giving birth to new areas of physics. Today, burgeoning fields such as High-Energy Particle Physics, Astrophysics, and Space Weather owe their existence to these early investigations into the cosmos.

## 2. Detection and spectrometry of Cosmic Ray species

Research at Mt. Aragats has been pivotal in cosmic ray physics, with distinct phases marking its history. The initial phase, the mass-spectrometric era lasting approximately 15 years, is detailed in Chilingarian et al. (2009a). During this foundational period, the Alikhanyan brothers utilized their mass spectrometer to measure charged particles' momentum and absorption length. This enabled effective mass analysis and identification of protons within cosmic ray flux (Alikhanyan et al., 1945). This method also provided the earliest indications of particles with masses between the muon and the proton. Although not all peaks observed in the mass distributions were later verified as genuine particles—only a select few were recognized as K and π-mesons—the discourse on intermediate-mass particles, termed 'varitrons' (Alikhanyan et al., 1951), spurred numerous significant experimental and theoretical studies. This debate catalyzed global interest in elementary particles, establishing the Aragats station as a critical hub for cosmic ray research. It is important to note that discerning genuine particle peaks within one- and two-dimensional distributions continues to be a paramount yet challenging task in High-Energy Physics. Despite advanced mathematical techniques, current efforts are not immune to errors, occasionally leading to premature announcements such as the 'discovery' of the pentaquark (Seife, 2004).

The years from 1958 to 1970 marked significant advancements in the calorimetric measurement of cosmic rays (CR). In 1958, Naum Grigorov and his team from the Institute of Nuclear Physics at Moscow State University, along with physicists from the Yerevan Physics Institute, installed the pioneering ionization calorimeter at the Aragats station (Grigorov et al., 1958). Their work illuminated the energy dependence of hadron-nuclei inelastic cross-sections, a finding later corroborated by direct measurements on Proton satellites and through accelerator experiments (Grigorov et al., 1970).

Further enhancements came in 1968-69 with the execution of two pivotal experiments: PION, led by Vahram Avakyan (Alikhanyan et al., 1974), measured the vertical fluxes of cosmic ray hadrons, while MUON, led by Tina Asatiani (Asatiani et al., 1980), focused on horizontal muons. PION was distinguished by first incorporating a transition radiation detector for particle identification, led by Albert Oganesian. The setup facilitated the measurement of the cross sections of proton, pion, and neutron inelastic interactions with lead and carbon nuclei at energies between 0.5 and 5.0 TeV. The muon magnetic spectrometer was outfitted with coordinate measuring systems, including wire and wide-gap spark chambers, enhancing the muon momentum measurement to 2.5 TeV. These experiments were among the first to employ Soviet computers, M220, and Armenia's minicomputer, NAIRI-2, for data collection and analysis.

The 1970s ushered in solar physics research at Aragats by installing neutron monitors type 18HM64 at both the Aragats and Nor-Amberd research stations under the guidance of Khachik Babayan. These installations laid the foundation for Aragats' status as a premier cosmic ray monitoring center in its 'new history.

### 3. Galactic Cosmic Ray research

In the 1980s, the realization that comprehensive CR studies required large detectors capable of measuring various CR species across extensive energy ranges led to the design of the ANI experiment. Conceptualized by YerPhI (Erik Mamijanyan) and the Lebedev Physics Institute of the USSR Academy of Sciences (Sergey Nikolsky), the ANI experiment aimed to detect Extensive Air Showers (EASs, Auger et al., 1939) in all detail, including electromagnetic and hadron components, as well as energies of muons. The envisaged setup included a large hadron calorimeter with a surface area of 1600 m², a 40 m² area underground magnetic spectrometer capable of detecting momenta up to 10 TeV over, and 200 m² of scintillators beneath 15 meters of soil and concrete to detect muons with energies above 5 GeV. Although the full ANI complex was not completed due to the dissolution of the USSR, two surface arrays emerged as noteworthy contributions to high-energy CR physics: MAKET-ANI (Fig.2), under the leadership of Vitaly Romakhin and Gagik Hovsepyan and GAMMA, led by Romen Martirosov (Fig. 3, Garyaka et al., 2002).

Launched in 1997, the MAKET-ANI surface array was operational in its complete form with approximately 100 plastic scintillators, each with an area of 1 m². The array's design allowed for a highly efficient selection of EAS cores from an area of about 3000 m² surrounding the array's geometric center, with an efficiency exceeding 95% for EASs initiated by primary particles with energies of $5 \times 10^{14}$ eV or more. This compact array, comprises continuously calibrated detectors, was aptly suited for precise energy and CR composition measurements at the 'knee' of the cosmic ray spectrum. Over a million EASs observed between 1999 and 2002 have been meticulously analyzed, contributing significantly to the estimation of energy spectra of light and heavy nuclei groups within the energy range of $5*10^{14}$ to $10^{17}$ eV.

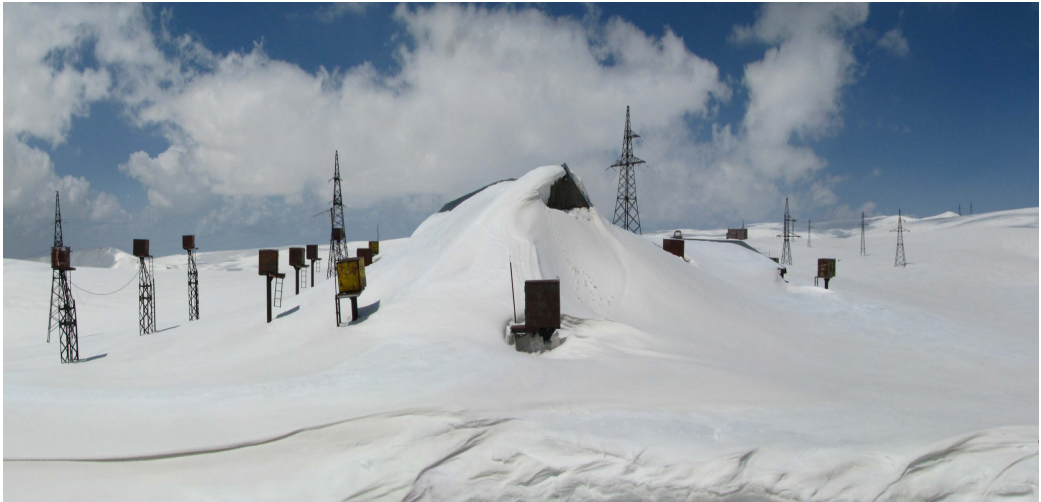

**Figure 2. MAKET-ANI detector researching physics around the "knee", first observation of the light and heavy Galactic nuclei energy spectra.**

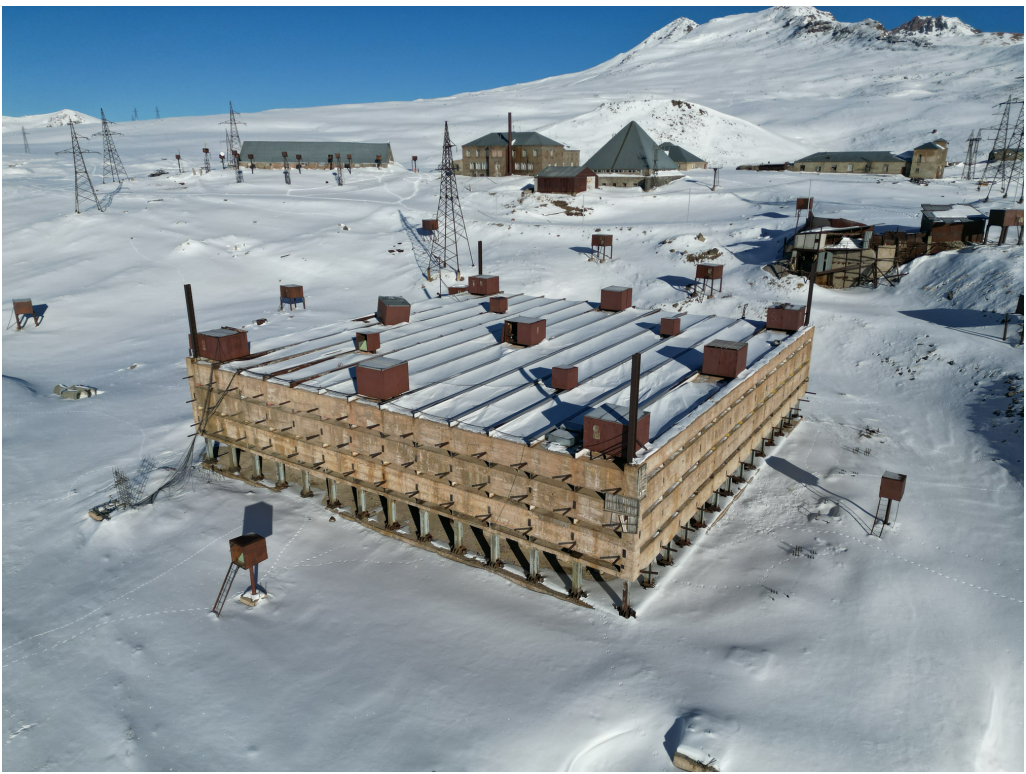

**Figure 3. The GAMMA detector of the ANI experiment registers "muon-poor" showers for selecting PeV energy gamma rays.**

In the 1980s, a groundbreaking AI-based nonparametric multivariate data analysis methodology was developed, tackling the CR inverse problem to deduce primary particle types and energies from diffuse, smeared EAS data (Chilingarian, 1989; Chilingarian and Zazyan, 1989; Chilingarian, 1994). This innovative approach enabled the derivation of partial energy spectra for distinct nuclear groups — light, intermediate, and heavy — and was first implemented in the analysis of data from the MAKET-ANI experiment (Chilingarian et al., 2004) and later applied to the extensive KASCADE experiment (Antony et al., 2002). Figure 4a presents a 2-way classification of the primary nucleus by the neural network into light and heavy classes, while Figure 4b encapsulates the physical findings of the MAKET-ANI experiment (Chilingarian et al., 2007). The data from MAKET-ANI reveal a pronounced 'knee' in the energy spectrum of the light component, consisting of protons and alpha particles, and a lack of such a feature in the heavy component spectrum up to approximately 30 PeV. These findings, highlighting a rigidity-dependent knee position, align with SNRs as probable sources of galactic cosmic rays and suggest Fermi-type acceleration as the prevailing mechanism for hadron acceleration.

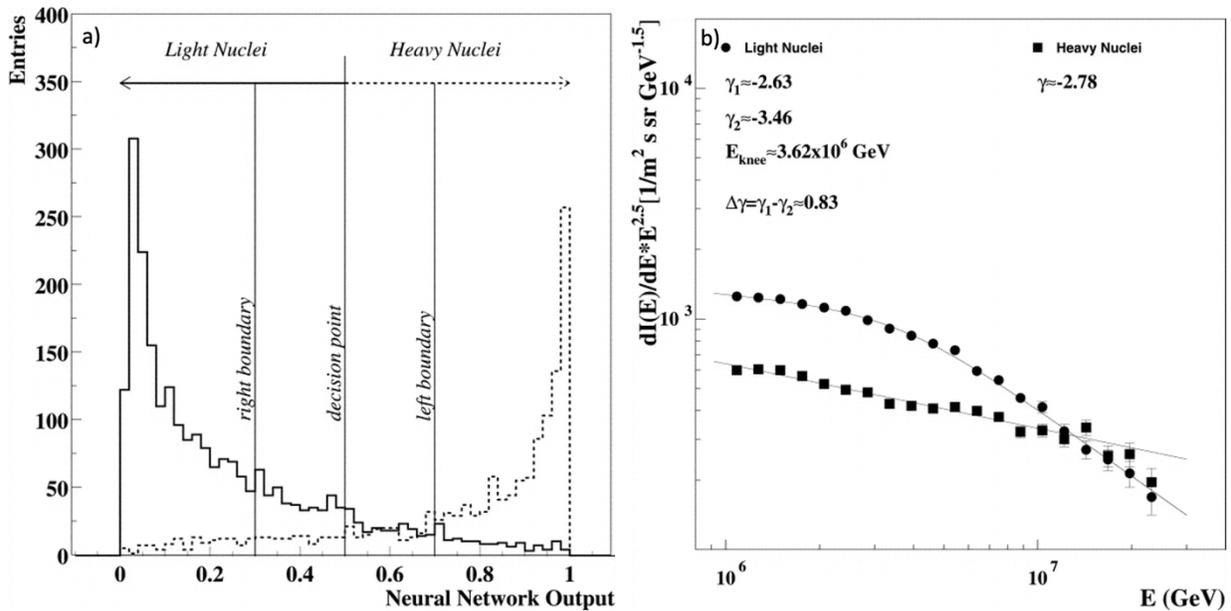

**Figure 4. a) The output of the Neural Network trained to distinguish "light" and "heavy" nuclei (from Chilingarian et al., 2004); b) Energy spectra of light and heavy nuclei obtained by neural classification and energy estimation. The EAS characteristics used are shower size and shape (age parameter)**

4. Solar Physics and Space Weather Research

The Aragats Space Environmental Center (ASEC), established in 2000, positions itself at the vanguard of solar physics and space weather research (Chilingarian et al., 2003; 2005). Utilizing neutron monitors and scintillation detectors from the MAKET-ANI and GAMMA arrays, ASEC has been instrumental in tracking CR fluxes. Its capabilities were notably demonstrated during the peak of the 23rd solar activity cycle between 2000 and 2003, precisely capturing numerous ground-level enhancements (GLEs) and Forbush decreases (FDs) (Chilingarian and Bostanjyan, 2009; 2010).

In 2001, ASEC pioneered the first early warning alert system against extreme solar energetic particle (SEP) events, characterized by hard spectra that pose risks to satellite electronics and the health of the Space Station crew (Gevorgyan et al., 2005). Figure 5 compares the Aragats neutron monitor's (ArNM) detection of the GLE and the arrival times of bulk 'hard' particles with energies exceeding 50 MeV, as recorded by GOES spacecraft detectors. The 'hard' particles, capable of penetrating spacecraft walls, can deliver significant radiation doses to astronauts and damage unmanned spacecraft electronics. They can also cause ionospheric disturbances, affecting radio communications. The data presented in Figure 5 juxtaposes ArNM count rates (left Y-axis) against the 'hard' (>50 MeV) solar proton fluxes detected by GOES spectrometers (right Y-axis) for four significant SEP events of the 23rd solar cycle. For all events, the intensification of harmful 'hard' particles lags behind the neutron bursts registered by ArNM. This indicates that the most energetic solar ions triggering GLEs arrive 30 minutes before the peak intensity of >50 MeV solar protons. Detecting these early-arriving relativistic ions by GLE measurement serves as a forewarning of impending hazardous solar particle fluxes. By continuously synchronizing data from ASEC's solar monitors and analyzing correlations between different types of secondary cosmic rays, ASEC can issue timely alerts about dangerous radiation events. An e-mail alert system, established by Babayan et al. (2001), notifies subscribers within 5 minutes of a sudden increase in count rate, affording satellite operators the critical window needed to enact protective measures. Later, a methodology was developed allowing fast recovery of the energy spectra of GLE ions (Chilingarian and Reimers, 2008; Zazyan and Chilingarian, 2009) to metalize upcoming hazards of SEP events.

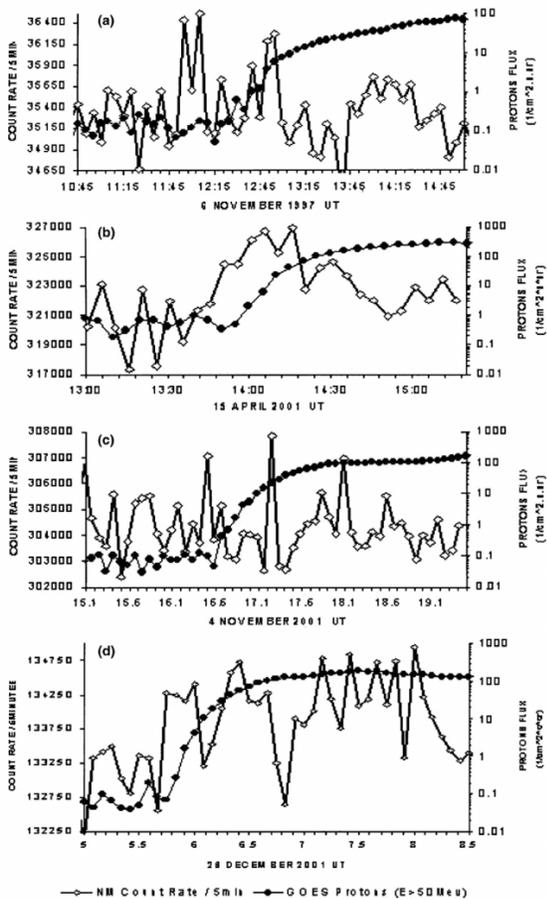

**Figure 5. Comparison of neutron count rates measured by ArNM and proton fluxes measured by facilities of GOES satellite for the most severe SEPS of the 23-rd solar activity cycle.**

Aragats Solar Neutron Telescope (ASNT) has been in operation at the Aragats research station since 1997 as part of a global network under the coordination of Nagoya University's Solar-Terrestrial Laboratory, led by Yasushi Muraki, seen in Fig. 6 near the ASNT assembly (Muraki et al., 1995). While ASNT's primary aim is to monitor direct neutron flux from the Sun, its capabilities extend to tracking charged fluxes and determining the trajectory of incoming muons.

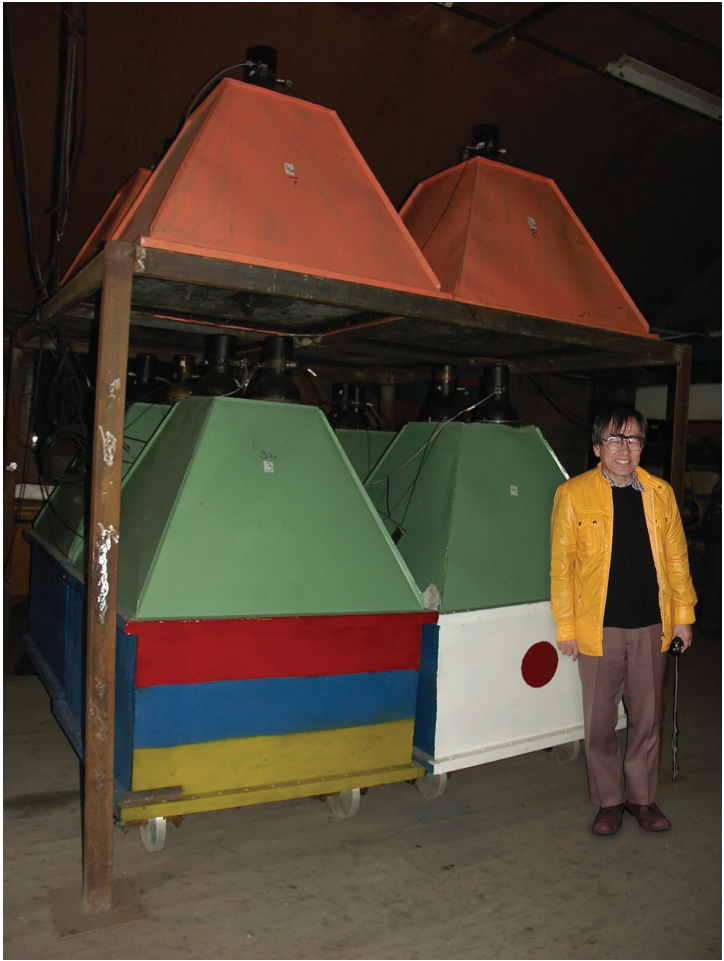

**Figure 6. Aragats Solar Neutron Telescope for registration of neutrons from violent solar bursts (worldwide network coordinated by Nagoya University, Professor Yasushi Muraki). The first measurement of the energy spectra of electrons from atmospheric accelerators was done in 2009 by ASNT.**

A landmark discovery at Mt. Aragats was the observation of solar protons with the highest energies ever recorded (Bostanjyan et al., 2007; Chilingarian, 2009c). On January 20, 2005, an intense X-ray burst erupted near the western edge of the Sun, with the X7.1 flare beginning at 06:36 UT and peaking at 07:01 UT. Just minutes after the flare's onset, the most rapid ground-level enhancement (GLE) of solar cycle 23 was detected. The GLE began at 06:48 UT, and the South Pole neutron monitor registered a staggering 5000% increase in intensity, the largest ever recorded by neutron monitors. Between 07:00 and 08:00 UT, ASEC detectors measured unusually high count rates. Subsequently, from 07:02 to 07:04 UT, the Aragats Multichannel Muon Monitor (AMMM, Fig. 7) marked the first detection of a significant increase (4σ) in the flux of muons greater than 5 GeV coinciding with a GLE. The differential energy spectra of solar protons showed a pronounced 'turn-over' at 700-800 MeV. Remaining exceptionally steep, the energy spectrum exhibited a power index of about -1 up to 800 MeV and extended into the tens of GeV range with a power index between -5 and -6. A meticulous statistical analysis of the peak suggested that the solar energetic event included protons with energies exceeding 20 GeV.

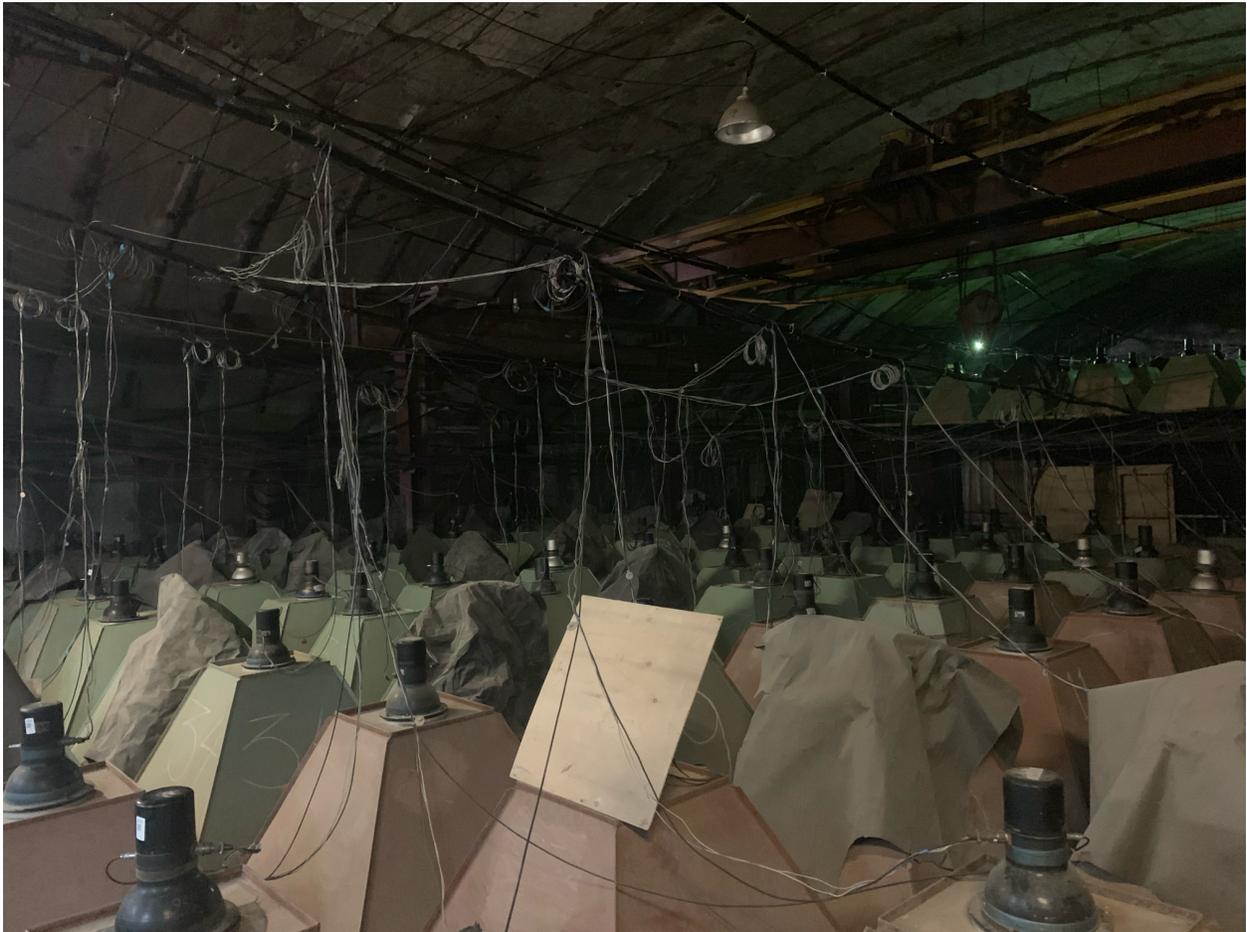

**Figure 7. On January 20, 2005, the GAMMA experiment's AMMM muon detector measured the ever-largest energy proton flux from solar accelerators (> 20 GeV).**

In 2007, the CRD inaugurated the SEVAN detector network (Space Environment Viewing and Analysis Network, Chilingarian et al., 2009d), marking a significant advancement in global particle detection. This initiative, falling under the umbrella of the United Nations Basic Space Science (UNBSS) activities, was facilitated by the International Heliophysical Year 2007 and the UN Office for Outer Space Affairs, which spearheaded an instrument deployment program in developing nations. CRD experts developed a new class of hybrid detectors capable of measuring neutral and charged particles and their energy spectra. The SEVAN Network's primary goal is to monitor and measure the fluxes of various secondary cosmic ray species, effectively creating a comprehensive tool for investigating particle acceleration and propagation in the corona and in interplanetary space. The network's initial rollout featured installations in Croatia, Bulgaria, and India. Expansion continued with the installment of SEVAN detectors in Croatia,

Slovakia, Germany (Hamburg and Berlin), Czech Republic, and atop Zugspitze, Germany's highest peak, in 2023, illustrated in the map, Fig. 8.

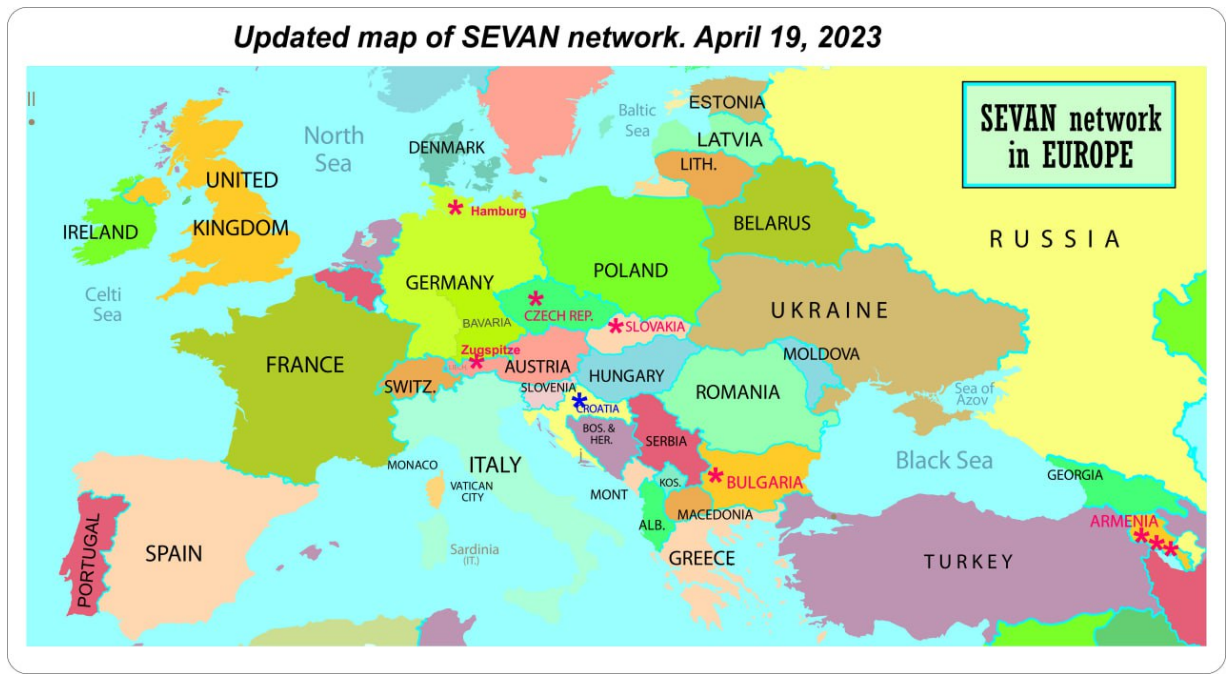

**Figure 8. Red asterisks show the hosts of the Space Environment Viewing and Analysis Network (SEVAN).**

CRD's advanced particle detector system has been underscored by the development of sophisticated data acquisition (DAQ) systems. Under the leadership of V. Danielyan and subsequently by Davit Pokhsraryan, the electronics group has designed and implemented versatile microcontroller-based DAQ systems. These systems are tailored to meet the rigorous demands for stability and efficiency within multi-channel, multi-detector setups. A network of hundreds of channels captures the dynamic fluxes of secondary particles, yielding critical data on solar modulation effects during major solar events (Arakelyan et al., 2016; Pokhsraryan, 2016). The deployment of advanced software for data analysis, triggering, and swift data transfer facilitates the aggregation of time series of cosmic ray flux count rates within the CRD's central databases (Chilingaryan et al., 2010). These databases serve as a reservoir for comprehensive multidimensional visualization and statistical analysis, proving indispensable tools for experimental data scrutiny, collaborative research, and scientific publication preparation.
Data integration from local and international detector networks into the MySQL database at CRD headquarters in Yerevan is streamlined through the ADEI platform, which offers advanced visualization and statistical analysis capabilities. ADEI empowers users to rapidly analyze data, craft figures and presentations, engage in collaborative data analysis with remote teams, test scientific hypotheses, and derive physical interpretations.
Moreover, real-time monitoring of thunderstorm events is made feasible through alert systems that dispatch notifications via email. The ADEI database encompasses a time series of both neutral and charged particle count rates, along with data on Near Surface Electric Field (NSEF) disturbances, measured by an array of Boltek EFM-100 electric field mills and meteorological conditions recorded by automatic weather stations from Davis Instruments. Consolidating these diverse data sets enables users to visualize and perform multivariate correlation analyses between particle fluxes and various environmental factors.

ASEC upholds a longstanding educational tradition with its high-energy physics schools, an initiative started by A. Alikhanyan in the 1960s at Nor Amberd. The SEE-2005 (Solar Extreme Events) and FORGES-2008 (Forecasting Of Radiation and Geomagnetic Storms) conferences were notable events that brought together a diverse group of scientists and students from various countries. These gatherings continue to foster collaboration and learning in the dynamic fields of solar physics and space weather research.

## 5. High-energy physics in Atmosphere (HEPA)

Following the most intense GLE event in 2005, meticulously documented by ASEC's detectors, solar activity transitioned into the notably tranquil 24th cycle. Concurrently, ASEC shifted its focus towards planetary science, spearheading the nascent High Energy Physics in Atmosphere (HEPA) field. HEPA explores cosmic ray flux modulation in Earth's atmosphere, including phenomena such as thunderstorm ground enhancements (TGEs). These TGEs, characterized by a sudden increase in cosmic ray electrons and gamma-rays during thunderstorms, have illuminated the intricate interplay between atmospheric processes and cosmic ray fluxes, as well as the influence of cosmic rays on our planet (Chilingarian et al., 2010; 2011).

The consensus among researchers is that TGEs are driven by the relativistic runaway electron avalanche (RREA) mechanism, prevalent in electrically charged atmospheric conditions (Gurevich et al., 1992; Babich et al., 2001; Alexeenko et al., 2002; Dwyer et al., 2003). This research has led to a significant expansion of the experimental infrastructure on Mt. Aragats, where a suite of new particle detectors, weather stations, lightning detectors, and electric field sensors have been deployed. The installation includes a network of seven NaI spectrometers to gather huge statistics for gamma-ray energy spectrometry within the 0.3 to 50 MeV range. Additionally, the site boasts three STAND1 detectors—each comprising three 1 cm thick, 1 m$^2$ area scintillators stacked atop each and a separate 3 cm thick standalone detector, surveying a 50,000 m$^2$ area. These detectors are integrated into a rapid data synchronization system capable of capturing time series with a 50 ms sampling rate, precisely aligned with atmospheric discharges at nanosecond precision. Sixteen plastic scintillators of the MAKET-ANI surface array continue to record EASs and electron avalanches originating from overhead thunderclouds. The ASNT remains a cornerstone in high-energy atmospheric physics, measuring the TGE electron energy spectrum in the 15 - 50 MeV energy band (Chilingarian et al., 2023).

Thunderstorms generate extensive electric fields that permeate the areas within and surrounding the storm system. These fields, depicted by red arrows in Fig. 9, arise from charge separations within the clouds due to warm air updrafts and the dynamic interaction of different hydrometeors, forming oppositely charged dipoles.

Pioneering research by Joachim Kuettner between 1945 and 1949 at Zugspitze (Kuettner, 1950) unveiled the intricate tripole charge structure within thunderclouds. According to the tripole model, atmospheric electric fields consist of upper and lower dipoles that drive free electrons toward space and the Earth's surface. The lower dipole is characterized by a central negative layer in the cloud, mirrored by a corresponding charge in the Earth. A third dipole forms between this layer and a transient lower positive charge region (LPCR)—associated with descending graupel (snow pellets coated with a layer of ice). Electrons propelled by lower dipole can initiate electron-gamma ray avalanches, observed on the ground as TGEs, comprising vast numbers of gamma rays and electrons, with the occasional production of neutrons.

Moreover, a fourth dipole between the lower positive charge region and its Earthly counterpart accelerates positrons and positive muons while slowing electrons and negative muons. A fifth dipole between the main positive layer and an upper screening layer propels electrons downwards, contributing to the complex interactions during a thunderstorm.

Balloon experiments in New Mexico (Marshall et al., 1995; Stolzenburg et al., 2007) and TGEs detected on Aragats (Chilingarian et al., 2019, 2022), Zugspitze (Chilingarian et al., 2024a), and Lomnicky Stit mountains (Chum et al., 2020) have underscored the relationship between particle fluxes and atmospheric electric fields. Observations confirm that strong electric fields above particle detectors facilitate the development of RREA (Chilingarian et al., 2023). Simulations using CORSIKA and GEANT4 codes support these findings, indicating a significant increase in particle numbers when the electric field exceeds critical thresholds.

Lightning flashes discharge the lower dipole below the RREA initiation threshold. This decrease leads to a weakening of RREA by eliminating high-energy particles. Nonetheless, even after the near-surface electric field strength returns to fair-weather value, the TGEs can continue due to the Radon circulation effect (Chilingarian et al., 2020). This is due to the presence of non-stable Radon chain isotopes $^{214}$Pb and $^{214}$Bi, which are lifted into the atmosphere and enlarge natural gamma radiation (see the central part of Fig. 9). TGE is prolonged up to 2-2.5 hours for energies below 3 MeV due to $^{214}$Pb and $^{214}$Bi isotopes with half-lives of approximately 27 and 20 minutes, respectively.

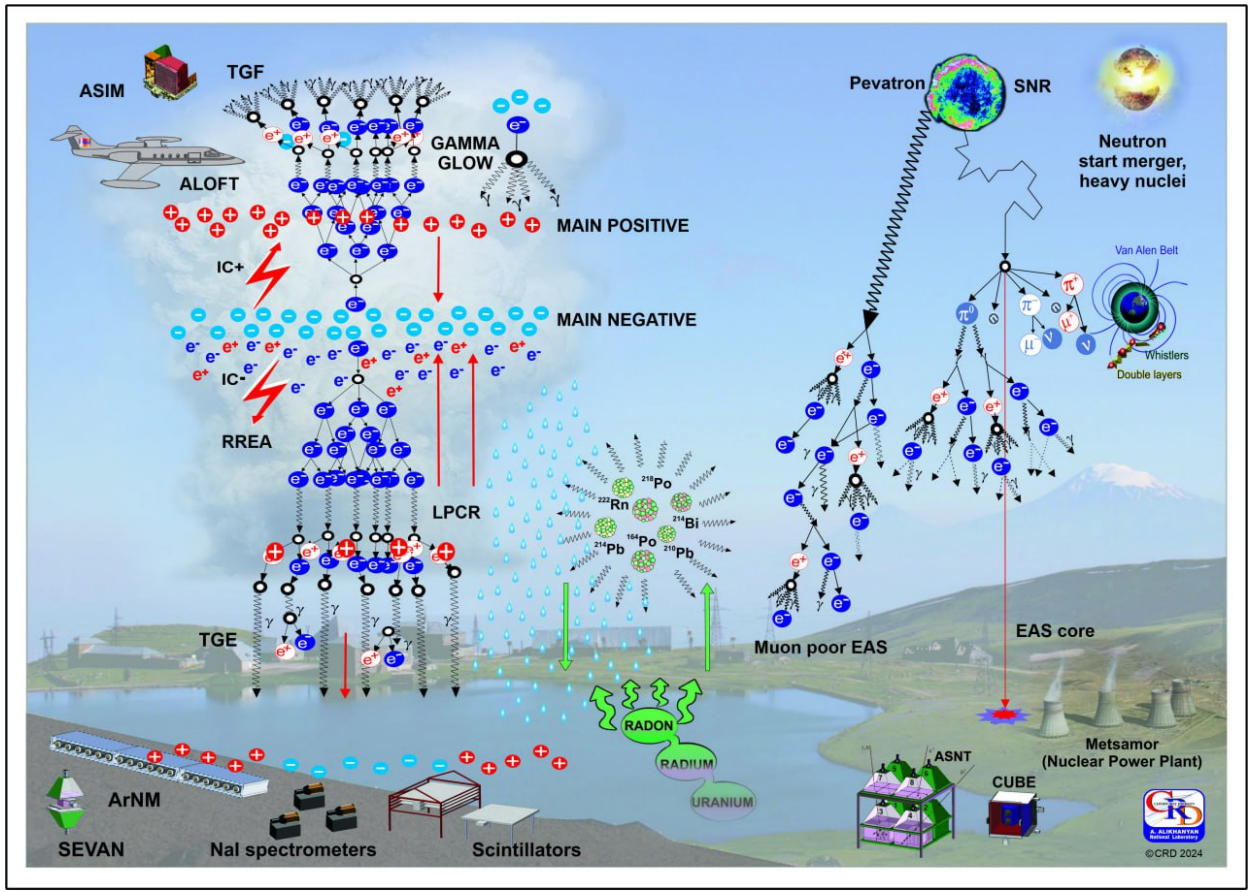

**Figure 9. The fluxes of secondary particles from space and atmospheric accelerators, as well as gamma radiation from $^{222}$Rn progeny, are shown. The cartoon also shows the charge structure of a thundercloud, the direction of electric fields, lightning flashes, and various measuring facilities on the surface and in space. Additionally, we can see the sources of primary cosmic rays. In the background is the Aragats cosmic ray station, located at 3200 meters and equipped with various particle detectors and spectrometers.**

The diagram on the right illustrates the EASs, which originate in the atmosphere when primary gamma rays and protons collide with atmospheric atoms. These high-energy particles are produced during cataclysmic cosmic events, such as supernova explosions and neutron star mergers within our Galaxy and beyond. While the field explores other sources of ultra-high-energy particles, contemporary research is propelled by new expansive experiments, often situated at high altitudes, to detect EASs. When EASs reach the Earth's surface, they can span several square kilometers, with their cores—indicated by the red circle in the diagram—harboring the most energetic secondary particles that create brief, intense bursts. At altitudes exceeding 4000 meters, the observation of muon-poor EAS events suggests the presence of Pevatrons, celestial accelerators capable of propelling protons to energies up to $10^{15}$ eV. Additionally, the diagram references a nuclear power plant as a potential source of radioactive contamination and the Van Allen radiation belt, known to direct MeV electrons toward the Earth's surface.

Over the past decade and a half, physicists at the Cosmic Ray Division have significantly advanced HEPA, an interdisciplinary field at the crossroads of cosmic rays and atmospheric physics. The latest breakthroughs from Aragats, observed in 2023, exemplify the dynamic nature of this research. Previous data reveals that impulsive enhancements of particle flux on Aragats typically are 10-20%, with rare instances exceeding 100%. However, in 2023, 5 TGEs enhanced the 100% limit, with one showing an unprecedented enhancement of 1800% (Chilingarian et al., 2024b). 369 TGEs documented between 2013 and 2023 shown in Fig. 10 demonstrate a specific seasonal pattern: the majority, including all significant events, transpired during Spring and Autumn (accounting for 81%), coinciding with temperatures ranging between -2°C and +2°C. These seasons are characterized by low-lying clouds over the Aragats research station, as indicated by the yellow and green segments in the histogram of Figure 10. Meanwhile, approximately 12% of TGEs occurred in the Summer (denoted by the rose color, often overlapping with black), and 7% in Winter (represented by the black color).

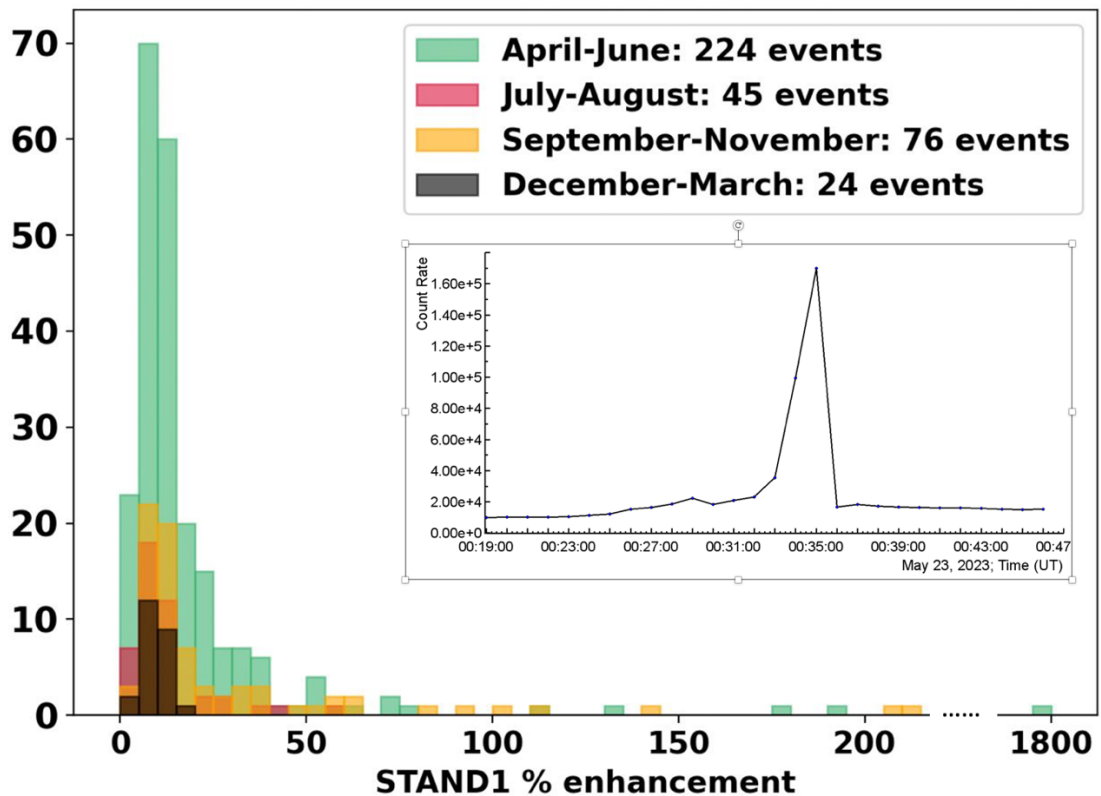

**Figure 10. The season-dependent histogram of the 369 TGE enhancements in percent. The 1 cm thick and 1 m² area plastic scintillator was used for the relative enhancement calculation. The inset shows the 1-minute time series of the scintillator count rate measured on May 23, 2023 (maximum flux was at 00:34 – 00:35).**

After an unusually tranquil 24th solar cycle, the Sun is exhibiting a marked increase in activity that heralds an intense 25th cycle, expected to peak in 2024. This uptick is anticipated to produce numerous violent coronal mass ejections that could impact Earth, leading to geomagnetic storms and SEP events. The complex interaction between the turbulent interplanetary magnetic fields and Earth's geomagnetic field can cause diverse phenomena, ranging from damaging satellite electronics to generating awe-inspiring Auroras. In this dynamic context, understanding the influence of large magnetized solar ejections on the near-Earth environment is crucial. Cosmic rays serve as messengers, carrying essential information about these multifaceted processes. Ground-based particle detector networks, which include the latest installations at Aragats and Zugspitze, offer valuable data that complement observations from space agencies like NOAA, NASA, and ESA.

On November 5, 2023, particle detector networks stationed at mid-latitude mountaintops observed a rare Magnetospheric Effect. The newly installed SEVAN detector at Zugspitze (Chilingarian et al., 2024a), see Fig. 11) and the CUBE detector on Aragats provide the first measurement of the energy spectrum of particles responsible for this magnetospheric phenomenon. With its significant role in verifying the RREA-TGE theory of particle bursts, the SEVAN network stands prepared for the solar maximum of 2024, ready to capture the energy spectra of imminent GLEs, FDs, and geomagnetic effects, thereby continuing the legacy of research conducted by the CRD during the 23rd solar cycle two decades ago.

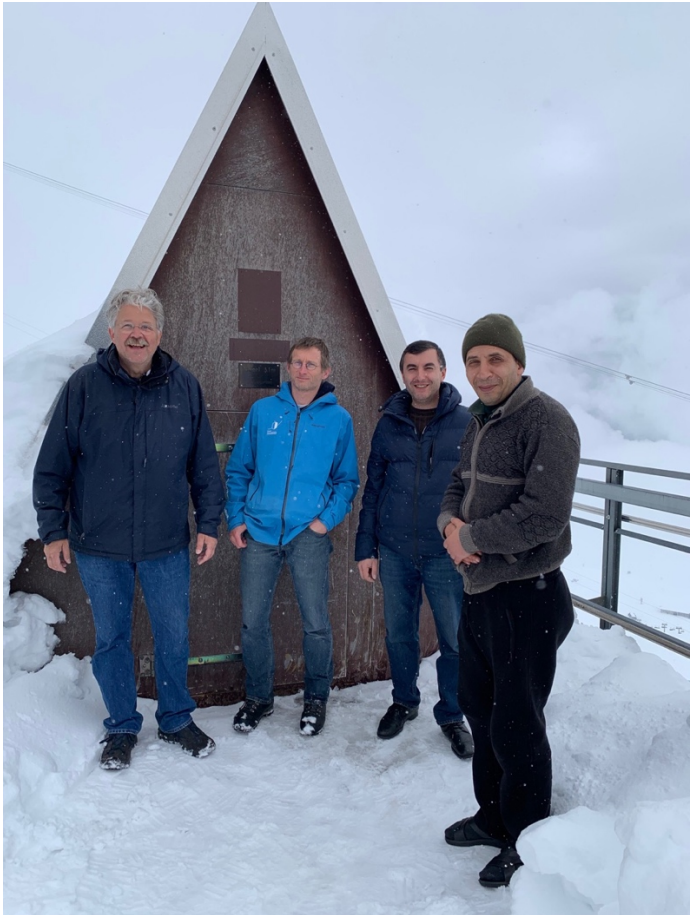

**Figure 11. Johannes Knapp, Till Rehm, Tigran Karapetyan, and Balabek Sargsyan after installing the SEVAN light detector at the Environmental research station Scheefernhaus (Zugspitze, 2650 m).**

On July 11, 2023, on Aragats, a remarkable 500% increase in positron flux was observed, coinciding with a significant TGE (Chilingarian et al., 2024c). The enhanced flux of electrons and gamma rays was attributed to RREA generated within the dipole formed between the main negatively charged layer in the middle of the thundercloud and the Lower Positively Charged Region (LPCR) at the bottom of the thundercloud. Concurrently, a substantial enhancement in the 511 keV gamma-ray flux resulting from electron-positron annihilation was recorded. This surge is intricately linked to the LPCR within the thundercloud. The emergence of the LPCR induces a polarity change in the atmospheric electric field (AEF) below the LPCR (fourth dipole), leading to the deceleration of electrons and the acceleration of positrons. Particle flux measurements were conducted using scintillation and NaI(TL) spectrometers. To mitigate the contamination of natural gamma radiation and refine the 511 keV flux measurements, the ORTEC spectrometer was shielded with a 4 cm thick lead filter. CORSIKA simulations corroborate the observed positron flux enhancement. Highlighting the synergy between high-energy physics in the atmosphere and astroparticle physics, we introduce a new scenario to elucidate the enigmatic large flux of galactic positrons measured by the Alpha Magnetic Spectrometer (AMS) aboard the International Space Station (ISS).

## 6. Conclusions

Delving into cosmic ray interactions with the Earth's atmosphere sheds light on essential atmospheric dynamics, offering a deeper understanding of the atmospheric modulation effects that lead to substantial particle bursts. The intricate relationship between cosmic rays and atmospheric electric fields is pivotal in identifying primary particles' types and energies. An integrated approach combining particle detection and atmospheric physics instrumentation has unraveled the factors contributing to EAS and RREA physics.

At the forefront of solar physics, space weather, and high-energy atmospheric physics, the CRD has established itself as a leading institution on a global scale. The SEVAN network, extending across the peaks of Armenia and Eastern Europe to Germany, is a versatile facility dedicated to comprehensive research in atmospheric and solar physics and space weather phenomena. The CRD's notable contributions to high-energy astrophysics encompass clarifying Galactic Cosmic Rays' origins and acceleration mechanisms and pinpointing the upper limits of solar proton acceleration, which exceed 20 GeV.

Recognition of the CRD's contributions is widespread, evidenced by over ten thousand citations in high-impact scientific journals, reflecting the division's profound influence and acclaim. The CRD's innovation extends beyond traditional research; it has pioneered cutting-edge AI-powered data analysis methodologies that facilitate multivariate correlations, with applications ranging from astroparticle physics to genomic studies.

The CRD's commitment to collaborative scientific progress is showcased through its active participation in international conferences and leadership in organizing the annual TEPA conferences at the Nor Amberd International Conference Center. These events are vital for advancing dialogue and partnership within the burgeoning field of high-energy atmospheric physics, which now includes solar physics.

In recent years, the station has continued to enrich scientific knowledge and evolved into a nexus for international scientific cooperation. The CRD invites researchers worldwide to utilize its laboratories for detector research and calibration, fostering a collaborative spirit. With anticipation, the scientific community looks to March for the announcement of international competition results, setting the stage for the subsequent TEPA meeting in October 2024.

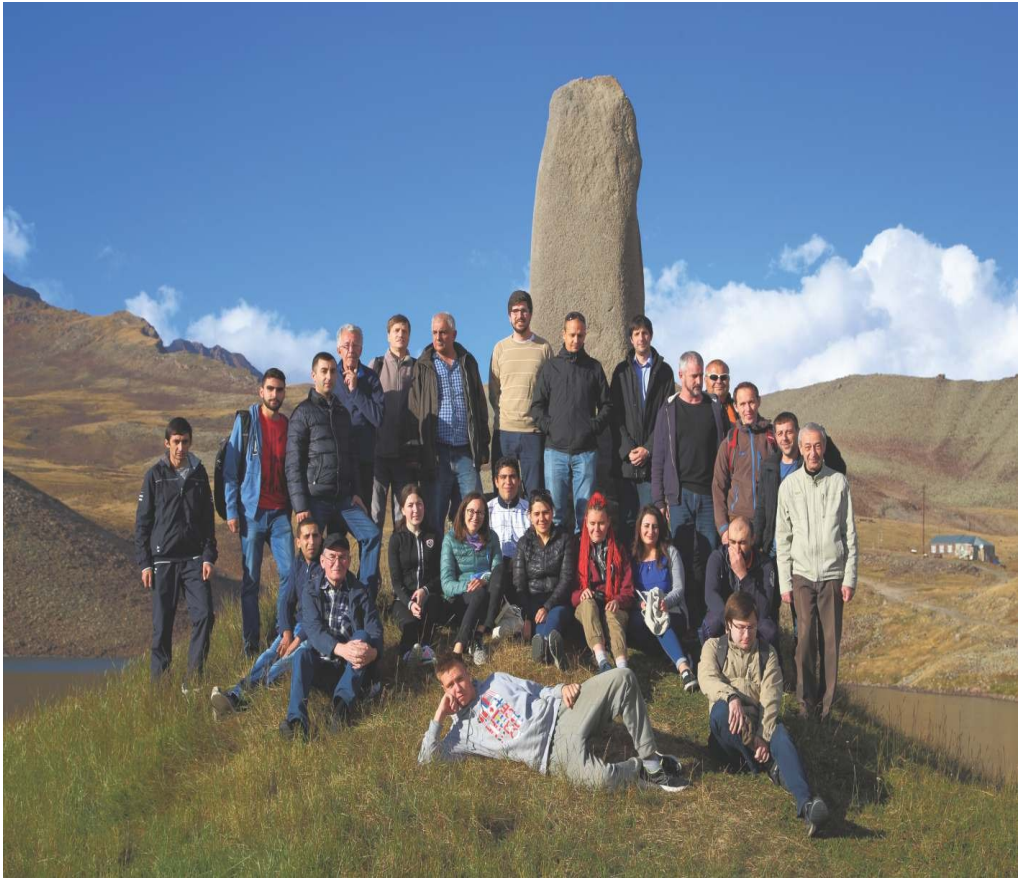

**Figure 12. TEPA -2019 (Thunderstorms and elementary particle acceleration) conference participants on Aragats**

**References**

Alexeenko V.V, Khaerdinov N.S, Lidvansky A.S, et al. (2002) Transient variations of secondary cosmic rays due to atmospheric electric field and evidence for pre-lightning particle acceleration. Phys Lett A 301:299–306.

Alikhanyan, A.I., Alikhanov, A.I., Nikitin, S. Highly ionizing particles in the soft component of cosmic rays. J. Phys. 9, 175–182, 1945.

Alikhanyan, A.I., Alikhanov, A.I. Varitrons. J. Exp. Theor. Phys. 21, 1023–1044 (in Russian), 1951.

Alikhanyan, A.I., Avakian, V.V., Mamidjanyan, E.A., et al. A facility for identification of the hadrons with energy 300 GeV with a transition radiation detector. Proc. Soviet Acad. Sci. Phys. Ser. 38, 1993–1995 (in Russian), 1974.

Antoni T., Apel W.D., Badea F. , et al., 2002. A non-parametric approach to infer the energy spectrum and the mass composition of cosmic rays, Astroparticle Physics 16 82002) 245-263.

Arakelyan, K., Abovyan, S., Avetisyan, A., et al. New electronics for the Aragats Space- Environmental Center (ASEC) particle detectors, in Chilingarian, A. (Ed.), Proceedings of International Symposium FORGES-2008, Nor Amberd, Armenia, TIGRAN METZ, pp. 105–116, 2009.

Asatiani, T.L., Chilingarian, A.A., Kazaryan, K., et al. Investigation of characteristics of high-energy cosmic ray muons. Proc. USSR Acad. Sci. Ser. Phys. 45, 323, 1980.

Auger P., Ehrenfest P., Maze, R., et al. (1939) Extensive Cosmic-Ray Showers, Reviews of Modern Physics, 11, 3–4, 288–291.

Babayan, V.Kh., Botsanjyan, N, Chilingarian, A. A, Alert service for extreme radiation storms, in Proceedings of the 27th International Cosmic Ray Conference, Hamburg, vol. 9, p. 3541, 2001.

Bostanjyan, N.Kh., Chilingarian, A.A., Karapetyan, G., et al. On the production of highest energy solar protons on 20 January 2005. Adv. Space Res. 39, 1456–1459, 2007.

Chilingarian, A.A. Statistical decisions under nonparametric a priory information. Comput. Phys. Commun. 54, 381–390, 1989.

Chilingaryan, A.A. Neural classification technique for background rejection in high energy physics experiments. Neurocomputing 6, 497–512, 1994.

Chilingarian, A.A., Avagyan, K., Babayan, V., et al. Aragats Space- Environmental Center: status and SEP forecasting possibilities. J. Phys. G Nucl. Part Phys. 29, 939–952, 2003.

A.Chilingarian , G. Gharagyozyan , G. Hovsepyan , S. Ghazaryan , L. Melkumyan , and A. Vardanyan,Light and Heavy CosmicRay Mass Group Energy Spectra as Measured by the MAKET-ANI Detector, Astrophysical Journal, 603, L29-L32, 2004, http://iopscience.iop.org/1538-4357/603/1/L29/fulltext/

Chilingarian, A.A., Arakelyan, K., Avagyan, K., et al. Correlated measurements of secondary cosmic ray fluxes by the Aragats Space Environmental Center monitors. NIM A 543, 483–496, 2005.

Chilingarian, A., Gharagyozyan, G., Hovsepyan, G., et al. Light and heavy cosmic-ray mass group energy spectra as measured by the MAKET-ANI detector. Astrophys. J. 603, L29–L32, 2004.

Chilingarian, A.A., Hovsepyan, G.G., Melkymyan, L.G., et al. Study of extensive air showers and primary energy spectra by MAKET-ANI detector on mountain Aragats. Astropart. Phys. 28, 58–71, 2007.

Chilingarian, A.A., Reymers, A. Investigations of the response of hybrid particle detectors for the Space Environmental Viewing and Analysis Network (SEVAN). Ann. Geophys. 26, 249–257, 2008.

Chilingarian A., Mirzoyan R., Zazyan M., Cosmic Ray research in Armenia, Advances in Space Research 44, 1183–1193, 2009a

Chilingarian, A., and Bostanjyan N. 2009b. Cosmic ray intensity increases detected by Aragats Space Environmental Center monitors during the 23rd solar activity cycle in correlation with geomagnetic storms, J. Geophys. Res., 114, A09107, doi:10.1029/2009JA014346.

Chilingarian, A.A. Statistical study of the detection of solar protons of highest energies on 20 January 2005. Adv. Space Res. 43, 702–707, doi:10.1016/j.asr.2008.10.005, 2009c.

Chilingarian, A., Avakyan, K., Arakelyan, K., et al. Space Environmental Viewing and Analysis Network (SEVAN) Earth, Moon, and Planets, doi 10.1007/s11038-008-9288-1, 2009d.

Chilingarian A., Bostanjyan, N. 2010. On the relation of the Forbush decreases detected by ASEC monitors during the 23rd solar activity cycle with ICME parameters, Advances in Space Research 45 (2010) 614.

Chilingarian A., Daryan A., Arakelyan K., Hovhannisyan A., Mailyan B., Melkumyan L., Hovsepyan G., and Vanyan L. (2010) Ground-based observations of thunderstorm-correlated fluxes of high-energy electrons, gamma rays, and neutrons (2010), Phys Rev D. 82, 043009.

Chilingarian A, Hovsepyan G., and Hovhannisyan A. (2011) Particle bursts from thunderclouds: Natural particle accelerators above our heads, Phys. Rev. D 83, 062001.

Chilingarian A, Mkrtchyan H., Karapetyan T., Chilingaryan S., Sargsyan B. and Arestakesyan A. (2019) Catalog of 2017 Thunderstorm Ground Enhancement (TGE) events observed on Aragats, Nature Scientific Reports 9(1):6253, DOI: 10.1038/s41598-019-42786-7.

Chilingarian, A., Hovsepyan, G., & Sargsyan, B. (2020). Circulation of Radon progeny in the terrestrial atmosphere during thunderstorms. Geophysical Research Letters, 47, e2020GL091155. https://doi. org/10.1029/2020GL091155.

Chilingarian A., Kozliner L., Sargsyan B., Soghomonyan S., Chilingaryan S., Pokhsraryan D., and Zazyan M. (2022) Thunderstorm Ground Enhancements: Multivariate analysis of 12 years of observations, Physical Review D 106, 082004.

Chilingarian, A., Hovsepyan, G., Aslanyan, D., Karapetyan, T., Sargsyan, B., & Zazyan, M. (2023) TGE electron energy spectra: Comment on "Radar diagnosis of the thundercloud electron accelerator" by E. Williams et al. (2022). *Journal of Geophysical Research: Atmospheres*, *128*, e2022JD037309, https://doi.org/10.1029/2022JD037309

Chilingarian A., Karapetyan T., Sargsyan B., Knapp J., Walter M., Rehm T., 2024a. Energy spectra of the first TGE observed on Zugspitze by the SEVAN light detector compared with the energetic TGE observed on Aragats, Astroparticle Physics 156, 02924

Chilingarian A., G. Hovsepyan, B. Sargsyan, T. Karapetyan, D. Aslanyan and L. Kozliner, 2024b. Enormous Impulsive Enhancement of particle fluxes observed on Aragats on May 23, 2023, Advances in Space Research, https://doi.org/10.1016/j.asr.2024.02.041
.
Chilingarian A., B. Sargsyan, Zazyan M. (2024c) An Enormous Increase in Atmospheric Positron Flux during a Summer Thunderstorm on Mount Aragats, Radiation Physics and Chemistry, in press.

Chilingaryan S., Beglarian A., Kopmann A., and Voekling S., Advanced data extraction infrastructure: A WEB based system for management of time series data, J. Phys. Conf. Ser. 219, 042034 (2010).

Chum, R. Langer, J. Baše, M. Kollárik, I. Strhárský, G. Diendorfer, J. Rusz (2020) Significant enhancements of secondary cosmic rays and electric field at high mountain peak during thunderstorms, Earth Planets Space 72, 28.

Danilova, T.V., Dunaevsky, A.M., Erlykin, A.D., et al. A project of the experiment on the investigation of interactions of hadrons in the energy range $10^3$–$10^5$ TeV. Proc. Armenian Acad. Sci. Phys. Ser. 17, 129–132 (in Russian), 1982.

Dwyer J.R. (2003) A fundamental limit on electric fields in air, Geophys. Res. Lett. 30, 2055.

Garyaka, A.P., Martirosov, R., Eganov, V., et al. The cosmic ray energy spectrum around the knee measured with the GAMMA array at Mt. Aragats. J. Phys. G. Nucl. Part. 28, 231–2328, 2002.

Gevorgyan, K. Avakyan, V. Babayan, et al., 2005. Test alert service against very large SEP Events, Advances in Space Research 36(12):2351-2356, DOI: 10.1016/j.asr.2004.04.016

Grigorov, N.L., Murzin, V.S., Rapoport, I.D. The method of measurement of the energy of particles in a region more than $10^{11}$ eV. J. Exp. Theor. Phys. 4, 506–523 (in Russian), 1958.

Grigorov, N.L., Nesterov, V.E., Rappoport, I.D. Measurements of particle spectra on the proton 1,2,3 satellites. J. Nucl. Phys. 11, 1058–1067 (in Russian), 1970.

Gurevich A. V., Milikh G. M., and Roussel-Dupre R. A. (1992) Runaway electron mechanism of air breakdown and preconditioning during a thunderstorm. Phys. Lett. 165A, 463.

Kocharian, N.M. Investigation of the azimuth asymmetry of cosmic rays, in Scientific studies of Yerevan State University, vol. 12, pp. 23–28, 1940 (in Russian).

Kocharian, N.M., Saakian, G.S., Aivazian, M.T. Energy spectra of l- mesons on the altitude of 3200 m. Rep. Armenian Acad. Sci. 24, 344– 348 (in Russian), 1957.

Kuettner, J. (1950) The electrical and meteorological conditions inside thunderclouds. *J. Meteorol.*, *7*, 322–332.

Marshall T.C, McCarthy M.P., and Rust W. D. (1995) Electric field magnitudes and lightning initiation in thunderstorms, J. Geophys. Res. 100, 7097.

Y. Muraki, S. Sakakibara, S. Shibata, M. Satoh, K. Murakami, T. Takahashi, K. R. Pyle, T. Sakai, and K. Mitsui, New Solar Neutron Detector and large Solar Flare Events of June 4th and 6th, 1991, J. Geomag. Geoelectr., 47 (1995) 1073-1078.

D. Pokhsraryan, Fast Data Acquisition system based on NI-myRIO board with GPS time stamping capabilities for atmospheric electricity research, in Proceedings of TEPA Symposium, Nor Amberd, vol. 2015, Tigran Mets, Yerevan, 2016, p. 23.

Seife, C. Rara avis or statistical mirage? Pentaquark remains at large. Science 306, 1281–1982, 2004.

Stolzenburg M., T., Marshall T.C., E. Bruning W.D., MacGorman D.R., and T. T. (2007) Electric ield values observed near lightning flash initiations, Geophys. Res. Lett., 34, L04804.

Zazyan M., Chilingarian, A. 2009. Calculations of the sensitivity of the particle detectors of ASEC and SEVAN networks to galactic and solar cosmic rays, Astroparticle Physics 32, 185.